# Impact of Extrinsic Interface Traps and Doping Atoms on Conductivity of Graphene Field Effect Devices

G I Zebrev, S A Shostachenko

**National Research Nuclear University MEPHI, Moscow, Russia**

E-mail: gizebrev@mephi.ru

**Abstract.** Near-interfacial oxide traps and chemical impurities on the graphene surface or at the graphene-dielectric interface can be a source of intentional or unintentional doping of graphene sheet. The efficiency of such chemical doping can vary in a wide range depending on parameters of graphene field effect devices. Mechanisms of such sensitivity of doping efficiency to the device characteristics need to be understood. The objective of this paper is to theoretically derive the analytical relations, adapted to the explicit calculation of graphene chemical doping.

## 1. Introduction

Graphene is a two-dimensional monolayer of carbon atoms arranged in a honeycomb hexagonal lattice [1]. Due to the unique structure of the lattice, graphene has a number of remarkable properties (zero bandgap, high carrier mobility etc.) which make it a promising candidate for advanced application in electronics. Despite significant progress in technology, the functional characteristics of graphene field effect device are still far from perfection. As a rule, graphene is not intrinsic material. Charged impurities in the dielectrics, or, chemical atoms or molecules on the surface of graphene sheet are able to dope the graphene, causing the charge neutrality voltage shifts and field-effect mobility and transconductance degradation. At the present, it is becoming increasingly clear that in order to fully describe the operation of graphene field-effect devices, one should control all reliability concerns associated with environmental charge trapping. We study theoretically the effects of external doping on characteristics of graphene field effect devices.

## 2. Doping and extrinsic doping of graphene

Chemical doping [2, 3] of graphene is an alternative to electrostatic doping, caused by the gate electrode controlled charging. The change in carrier concentration and conductivity of graphene can occur due to external defects, atoms, or chemical molecules. For example, the absorbed gas molecules acting as donors or acceptors. Graphene p-doping occurs even naturally for samples exposed to atmospheric water molecules [4].

Most frequently unintentional doping p-doping is observed, which is believed to originate from adsorbed water molecules due to the water molecules are acceptors. Removing the surface chemical impurities one can significantly reduce unintentional doping. For example, as reported in [5], the intrinsic doping level significantly lowers or vanishes under heating or when placing the graphene under vacuum and pumping for an extended time.

*2.1. Energy diagram*

The donor-like and acceptor-like surface impurity atoms and interface defect at graphene-oxide substrate can lead to the shift of the charge neutrality point (CNP) of the gate voltage $V_{NP}$. This effect quantitatively depends on interface trap density of energy states both for acceptor $D_{it}^{acc}$ and donor $D_{it}^{don}$ traps.

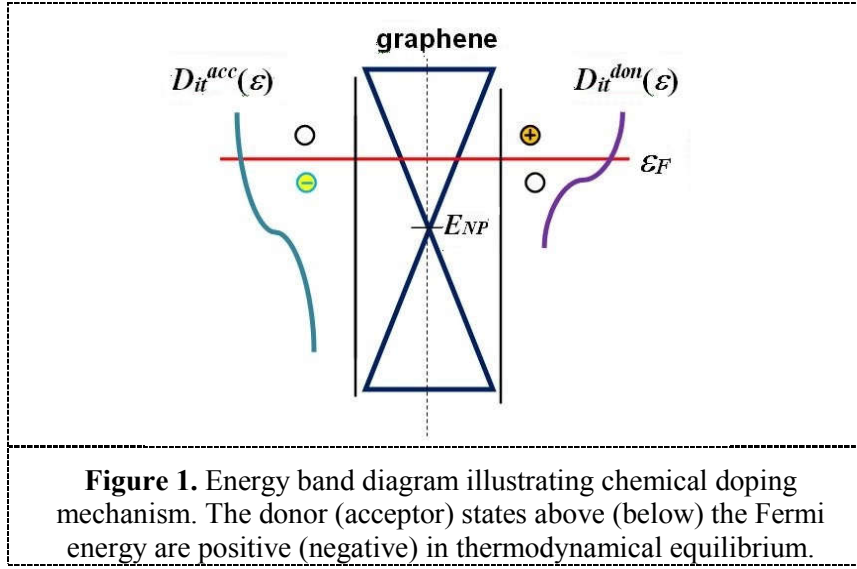

**Figure 1.** Energy band diagram illustrating chemical doping mechanism. The donor (acceptor) states above (below) the Fermi energy are positive (negative) in thermodynamical equilibrium.

Qualitative mechanism of chemical doping is shown in Figure 1. For example, the donor traps (typically, metal atoms) transfer own electrons to the graphene sheet, remaining in a positively charged state. Generally, the trap charge state and the direction of charge transfer depend on the graphene Fermi energy position with respect to the neutrality point. In its term, the Fermi energy is determined by density of charge traps. A self-consistent problem of computation of Fermi energy, neutrality point gate voltage $V_{NP}$ and charge concentration in graphene will be considered in the next section.

**3. Graphene charge densities and Fermi energy as function of gate voltage**

A self-consistent calculation of chemical doping effects requires an exact quantitative characterization of charge density and the Fermi energy in graphene. We will rely on this section on the analytical results reported in [6, 7, 8].

*3.1. Total and net charge densities in graphene*

The amount of charge in the two-dimensional graphene can be characterized by the two types of densities. First, it is the total carrier concentration $N_S$, which is a sum of the electron ($n_e$) and the hole ($n_h$) concentration

$$N_S = n_e + n_h \tag{1}$$

Second, it is the charge imbalance density (net charge density),

$$n_S = n_e - n_h . \tag{2}$$

The former is responsible for conductivity of graphene and its I-V characteristics. The latter controls electrostatics and C-V characteristics. Notice, the total charge density practically equals to the net charge density ($N_S \cong n_S$) excepting the vicinity of the charge neutrality point $\varepsilon_F > |k_B T|$.

*3.2. Fermi energy as function of gate voltage*

The Fermi energy $\varepsilon_F$ as function of gate voltage $V_G$ depends on the charge neutrality point gate voltage $V_{NP}$, and it can be written down as follows [6]

$$\varepsilon_F = \left(m^2\varepsilon_a^2 + 2\varepsilon_a e|V_G - V_{NP}|\right)^{1/2} - m\varepsilon_a = \frac{V_a}{m}\left[(1+2Y)^{1/2} - 1\right], \quad (3)$$

where

$$Y = \frac{|V_G - V_{NP}|}{V_a}, \quad (4)$$

and $V_a$ is defined as follows

$$V_a \equiv m^2 \varepsilon_a / e, \quad (5)$$

$\varepsilon_a$ is a characteristic graphene-oxide constant [6]

$$\varepsilon_a = \frac{\pi \hbar^2 v_0^2 C_{ox}}{2e^2}, \quad (6)$$

$C_{ox} = \varepsilon_{ox}\varepsilon_0 / d_{ox}$ is the gate oxide specific capacitance. The energy $\varepsilon_a$ varies in the range from ~1 meV at $d_{ox}$ ~ 200 nm and $\varepsilon_a$ =4 (SiO$_2$) to $\varepsilon_a$ ~ 0.5 eV at $d_{ox}$ ~2 nm and $\varepsilon_{ox}$ = 16 (HfO$_2$). A dimensionless "ideality" factor m in (3) and (5) is defined as

$$m = 1 + \frac{C_{it}}{C_{ox}} \quad (7)$$

where $C_{it}$ is an average interface trap capacitance, which is closely connected with energy density of interface traps $C_{it} = e^2 D_{it}$.

Total carrier concentration can be expressed as an exact relation [8]

$$N_S = n_e + n_h = \frac{\varepsilon_F^2}{\pi \hbar^2 v_0^2} + n_{res}, \quad (8)$$

$n_{res}$ is the residual carrier concentration at the CNP which assumed to be exactly equal to the intrinsic concentration $n_i$ in ideally homogeneous graphene or the total sum of carrier concentration in electron-hole puddles at the CNP. Notice, that the identities $N_S(-\varepsilon_F) = N_S(\varepsilon_F)$ and $n_S(-\varepsilon_F) = -n_S(\varepsilon_F)$ are valid immediately directly under definitions of this variables.

*3.1. Total charge as function of gate voltage*

Using (3) and (8) $\varepsilon_F(V_G)$ one obtains

$$e\delta N_S \equiv eN_S - en_{res} = C_{ox}\left(|V_G - V_{NP}| + V_a - V_a\left[1 + \frac{2|V_G - V_{NP}|}{V_a}\right]^{1/2}\right) = C_{ox}V_a\left[Y + 1 - (1+2Y)^{1/2}\right] \quad (9)$$

The vicinity around the CNP is characterized by an inequality $Y \ll 1$ ($|V_G - V_{NP}| < V_a$). The width of this region (~ $V_a$) is significantly dependent on the thickness of gate insulator and density of interface traps (see (5) and (7)). Given $Y \ll 1$, we have

$$\varepsilon_F \cong \frac{eV_a}{m}Y = e\frac{|V_G - V_{NP}|}{m}, \quad (10)$$

$$e\delta N_S \cong C_{ox}V_a \frac{Y^2}{2} = C_{ox}\frac{(V_G - V_{NP})^2}{2V_a} = \frac{e^2(V_G - V_{NP})^2}{\pi m^2 \hbar^2 v_0^2}, \quad (11)$$

Charge concentration in this region is dependent on the Plank constants, revealing (11) as a quantum result. This is explained by the quasi-classical approximation is failed near the charge neutrality point, where the carrier's de Broglie wavelength is not small. In contrast, given $Y \gg 1$ ($|V_G - V_{NP}| \gg V_a$), we have almost classical "electrostatic" result similar to that in silicon MOSFETs

$$e\delta N_S = C_{ox} V_a \left[ Y + 1 - (1 + 2Y)^{1/2} \right] \sim C_{ox} |V_G - V_{NP}|. \tag{12}$$

Despite the fact that the equation $e\delta N_S = C_{ox}|V_G - V_{NP}|$ is widely utilized as an accurate result, in reality, we always have an inequality $e\delta N_S < C_{ox}|V_G - V_{NP}|$.

## 4. Charge neutrality gate voltage and chemical doping effects

*4.1. Electrostatics equation*

The basic electrostatic equation in graphene field effect devices with non-uniform energy spectra of interface traps can be written down as follows [6]

$$V_G = \varphi_{GG} + \varepsilon_F + \frac{e^2 n_S}{C_{ox}} + \frac{e^2}{C_{ox}} \left[ \int_{-\infty}^{\varepsilon_F} D_{it}^{(acc)}(\varepsilon) d\varepsilon - \int_{\varepsilon_F}^{+\infty} D_{it}^{(don)}(\varepsilon) d\varepsilon \right] - \frac{Q_f}{C_{ox}}, \tag{13}$$

where, energy $\varepsilon$ is reckoned from the graphene CNP, $D_{it}^{(acc)}$ and $D_{it}^{(don)}$ are the acceptor and donor-like trap energy densities correspondingly, which assumed to be generally independent, $Q_f$ is density of fixed (i.e., graphene Fermi independent) charge trapped near the graphene – dielectric substrate interface, $\varphi_{GG} = W_G - W_{graphene}$ is the work function difference between graphene $W_G \cong 4.23\,\text{eV}$ and the gate material. It is explicitly assumed in (13) the occupied acceptors are negative and the unoccupied donors are positive. Equation (13) is valid both for electron and hole charging, i.e., for any sign of Fermi energy, including $\varepsilon_F = 0$. The charge neutrality point position is determined by the gate voltage corresponding to zero Fermi energy $V_{NP} = V_G(\varepsilon_F = 0)$

$$V_{NP} \equiv V_G(\varepsilon_F = 0) =$$
$$= \varphi_{GG} + \frac{e^2}{C_{ox}} \left[ \int_{-\infty}^{0} D_{it}^{(acc)}(\varepsilon) d\varepsilon - \int_{0}^{+\infty} D_{it}^{(don)}(\varepsilon) d\varepsilon \right] = \varphi_{GG} + \frac{Q_{acc}^{(-)}(\varepsilon_F = 0) + Q_{don}^{(+)}(\varepsilon_F = 0)}{C_{ox}} - \frac{Q_f}{C_{ox}}, \tag{14}$$

where $Q_{acc}^{(-)}(\varepsilon_F = 0) < 0$ and $Q_{don}^{(+)}(\varepsilon_F = 0) > 0$ are the charges of acceptor and donor traps at the neutrality point. Taking into account (14), the basic equation (13) can be rewritten as follows

$$e(V_G - V_{NP}) = \varepsilon_F + \frac{e^2 n_S}{C_{ox}} + \frac{e^2}{C_{ox}} \int_0^{\varepsilon_F} D_{it}(\varepsilon) d\varepsilon, \tag{15}$$

where the total interface trap is defined

$$D_{it}(\varepsilon) = D_{it}^{(acc)}(\varepsilon) + D_{it}^{(don)}(\varepsilon). \tag{16}$$

Notice it is rather hard in practice to distinguish the acceptor and the donor interface traps. Despite their opposite signs, these two types of traps electrically recharged in the same manner. Interface trap capacitance per unit area $C_{it}$ and interface trap level density $D_{it}(\varepsilon)$ are defined in the following way

$$C_{it}(\varepsilon_F) \equiv e \frac{d}{d\varepsilon_F} \left( Q_{acc}^{(-)}(\varepsilon_F) - Q_{don}^{(+)}(\varepsilon_F) \right) = e^2 D_{it}(\varepsilon_F) \tag{17}$$

where the total specific interface trap capacitance is a sum of acceptor and donor state capacitances.

*4.2. Graphene charge as function of chemical doping*

Let us calculate the density of graphene charge at $V_G = 0$ in a simple case of Fermi-energy independent oxide-trapped charge $Q_f$. This means that we set the interface trap charge (not density of states!) equals to zero at $V_G = 0$ or it is included in the fixed charge $Q_f$. In this case, we obtain the explicit equation for charge concentration in chemically doped graphene at the zero gate voltage

$$e|n_S| = C_{ox}\left(|V_{NP}| + V_a - V_a\left[1 + \frac{2|V_{NP}|}{V_a}\right]^{1/2}\right) = C_{ox}V_a\left[\left|\frac{C_{ox}V_{NP0} - Q_f}{C_{ox}V_a}\right| + 1 - \left(1 + 2\left|\frac{C_{ox}V_{NP0} - Q_f}{C_{ox}V_a}\right|\right)^{1/2}\right], \quad (18)$$

where $V_{NP0}$ is the CNP voltage position without any chemical doping. Figure 2 shows calculated results.

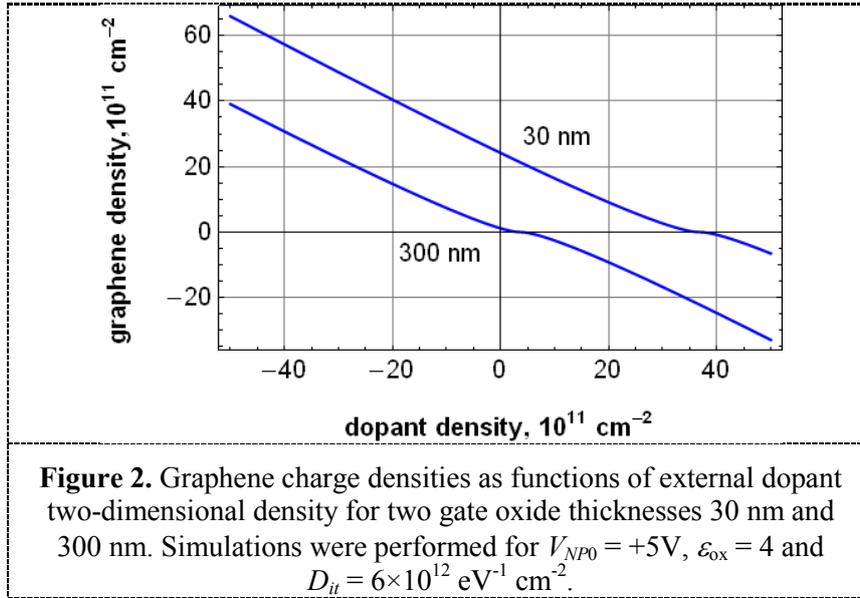

**Figure 2.** Graphene charge densities as functions of external dopant two-dimensional density for two gate oxide thicknesses 30 nm and 300 nm. Simulations were performed for $V_{NP0}$ = +5V, $\varepsilon_{ox}$ = 4 and $D_{it} = 6\times10^{12}$ eV$^{-1}$ cm$^{-2}$.

As can be seen in Fig. 2, the results of chemical doping depend strongly on capacitance (i.e., on the gate oxide thickness and dielectric constant) of gate insulators. Moreover, the efficiency of chemical doping is very sensitive to an initial value of $V_{NP0}$ and energy density of interface traps.

## 5. Conclusions

We have derived the explicit relations for calculation of carrier concentrations of graphene as functions external charge for different parameters of graphene field effect devices (gate insulator thickness and dielectric constants, interface trap energy density, initial values of neutrality point gate voltages).